# Decoupling effects of the resistive-switching behavior on the polarization reversal in ultrathin ferroelectric $Hf_{0.5}Zr_{0.5}O_2$ films

Chao Zhou[1,6], Sizhe Huang[1,6], Yangyang Si[1], Zhongqi Ren[1], Jianyuan Zhao[1], Hailin Wang[1], Jingxuan Li[1], Xianlong Cheng[1], Haoliang Huang[2], Shi Liu[3], Sujit Das[4], Shiqing Deng[5,*], Zuhuang Chen[1,7,*]

[1] State Key Laboratory of Precision Welding and Joining of Materials and Structures, School of Materials Science and Engineering, Harbin Institute of Technology, Shenzhen, 518055, China

[2] Quantum Science Center of Guangdong-Hong Kong-Macao Greater Bay Area, Shenzhen 518045, China

[3] Key Laboratory for Quantum Materials of Zhejiang Province, Department of Physics, School of Science, Westlake University, Hangzhou, Zhejiang 310024, China

[4] Material Research Centre, Indian Institute of Science, Bangalore 560012, India

[5] Beijing Advanced Innovation Center for Materials Genome Engineering, University of Science and Technology Beijing, Beijing, 100083, P.R. China

[6] These authors contributed equally to this work.

[7] Lead contact

*Correspondence: zuhuang@hit.edu.cn; sqdeng@ustb.edu.cn

**SUMMARY**

HfO$_2$-based ferroelectric films have attracted considerable attention as their nanoscale ferroelectricity and compatibility with CMOS technology, fulfilling demands of emerging memory technologies. However, as films scale down, resistive-switching behavior becomes increasingly pronounced, intricately intertwining with the polarization-switching process and affecting ferroelectric switching—factors often overlooked yet crucial for device performance optimization. By characterizing resistive-switching behavior and oxygen vacancy motion using tailored electric-pulse schemes, we decouple the resistive-switching behavior from the overall switching process in ultrathin ferroelectric Hf$_{0.5}$Zr$_{0.5}$O$_2$ films, which would otherwise erroneously inflate polarization values and increase coercive fields. Building on this, we elucidate endurance degradation mechanisms from dual perspectives of resistive switching and defect migration. Furthermore, we demonstrate the mitigated resistive-switching activity by designing HfO$_2$-based devices with symmetric oxide electrodes, achieving reduced coercive fields and improved cycling performances. This work provides crucial insights into the origins of inflated polarizations and reliability challenges in HfO$_2$-based devices while offering a viable strategy to enhance ferroelectric properties for advanced memory applications.

**INTRODUCTION**

The ferroelectricity found in the $HfO_2$-based thin films, a commercialized high-$κ$ gate material,[1] has been marked as a milestone for the development of low-power-consumption ferroelectric memories.[2] The superior CMOS compatibility and scalability embodied in the $HfO_2$-based ferroelectrics overcome the integration bottleneck of conventional ferroelectric materials and greatly facilitate the progress of emerging memory paradigms like in-memory computing and others.[3] However, as research advances, it has become evident that the electrical properties and phase structure of $HfO_2$-based films are intricately related to oxygen vacancies ($V_O$).[4–6] $V_O$ represent a predominant type of defect in $HfO_2$ materials, characterized by the prevalent oxygen-deficient compositions within $HfO_2$-based films.[7,8] Both the distribution condition and migration feature of $V_O$ would significantly influence electrical behaviors of $HfO_2$-based devices.[9–11] The drift of $V_O$ under electric fields imparts memristive attributes to $HfO_2$-based devices.[10,12,13] Meanwhile, the introduction of $V_O$ is found to be helpful for the stabilization of metastable polar phase.[5,14,15] Nevertheless, the electric field-driven $V_O$ redistribution, homogenization and migration,[6,16–18] in conjunction with $V_O$-related phase transformations,[4,19,20] can also give rise to wake-up, split-up, fatigue and breakdown phenomena of $HfO_2$-based ferroelectric films. Furthermore, unlike conventional ferroelectrics, $HfO_2$-based ferroelectric devices exhibit atypical switching characteristics, such as an anomalous increase in polarization values with rising temperatures[21] and extraordinarily divergent reported polarization values across different literatures (5 $\mu C/cm^2$ – 387 $\mu C/cm^2$).[6,22–25] These atypical behaviors suggest a strong influence of $V_O$ migration on the ferroelectric response, highlighting the critical role of defect motion in shaping the performance and reliability of $HfO_2$-based ferroelectric devices.

With the growing demand for miniaturized, high-density, and highly integrated memory devices,[26,27] ultrathin $HfO_2$-based ferroelectrics are garnering increasing interest owing to their unique size effect, which is exemplified by their sustained ferroelectricity in reduced film thicknesses.[28] Nevertheless, the synergy of ultrathin dimensions and the inherently large coercive field ($E_c$) in $HfO_2$-based ferroelectric devices facilitates a much more obvious oxygen ion/vacancy migration,[6,8] simultaneously aggravating the resistive-switching phenomenon. Therefore, the coupling between polarization-switching and resistive-switching (hereafter CPR in short) emerges as a prevalent yet often overlooked issue when it comes to the ultrathin $HfO_2$-based ferroelectric applications. Despite that, discussions about CPR are quite rare, and effects of CPR on ferroelectric switching characteristics have not been thoroughly studied either.

In this study, ultrathin $Hf_{0.5}Zr_{0.5}O_2$ (HZO) ferroelectric films serve as a model system to systematically investigate the interplay between resistive switching and polarization switching. By analyzing resistive-switching feature and oxygen defect migration behavior within the Pt/HZO/$La_{0.67}Sr_{0.33}MnO_3$ (LSMO) metal-ferroelectric-metal (MFM) capacitors, we successfully decouple the resistive switching behavior from the overall switching process, clarifying the origin of spurious "large" polarization values as well as increased $E_c$ frequently observed in ultrathin $HfO_2$-based devices—both of which are often misinterpreted in ferroelectric characterization. Based on these findings, we further reveal fatigue and endurance degradation mechanisms in HZO-based devices from the oxygen defect drift view. Leveraging the characteristics of defect migration, a "dielectric training" process is introduced to suppress the resistive switching effects, thereby enhancing polarization switching efficiency. Furthermore, an ultrathin HZO-based ferroelectric capacitor with symmetric LSMO oxide electrodes is fabricated, effectively mitigating the adverse effects of the CPR phenomenon. This approach leads to improved polarization stability and enhanced cycling performance, demonstrating a viable strategy to optimize the reliability of HZO-based ferroelectric devices.

**RESULTS**

**The structure and atypical electric behaviors of the ultrathin HZO device**

~6-nm-thick ultrathin HZO films were synthesized on (001)-oriented LSMO-buffered $SrTiO_3$ (STO) single crystal substrates using pulsed laser deposition (PLD) technique (details provided in the Methods Section). The thickness of the as-grown LSMO/HZO heterostructure was determined by X-ray reflection (XRR) and refined using the Nelder Mead algorithm (Figure S1a), which employs a least squares method. The structural properties of the heterostructures were characterized using $2\theta$-$\omega$ X-ray diffraction (XRD) (Figure S1b). The diffraction peak located at 29.94° corresponds to the (111) reflection of ferroelectric orthorhombic phase of HZO, with a $d$-spacing of 2.98 Å. The pronounced oscillations surrounding the diffraction peak for the ultrathin film indicates high crystalline quality and well-defined interfaces. Additionally, a weak peak observed at 34.2° is attributed to the (002) reflection of the paraelectric phase, potentially indicative of monoclinic or tetragonal structure.[29,30]

Despite the high crystalline quality of the ultrathin films, Pt/HZO/LSMO ferroelectric devices based on the 6 nm HZO film (Figure S1c) exhibit significant electric-response deviations compared to capacitors with thicker HZO layers. As depicted in Figure 1a, elevated fields were gradually applied to the 6 nm ultrathin HZO device. Interestingly, despite being measured below coercive field ($E_c$~ 4 MV/cm), the nominal polarization-electric field ($P$-$E$) loops of the ferroelectric device (Figure 1a) exhibit typical lossy dielectric characteristics. The accompanying current response (Figure S2a) reveals a notable leakage current superimposed on dielectric contribution, contributing to the observed "fat" $P$-$E$ loops. Quantitative analysis of the leakage current density (Figure 1b and Figure S2b) further confirms that the 6 nm HZO film demonstrates substantially higher leakage than conventional dielectric materials.[31] Notably, the ultrathin Pt/HZO/LSMO device manifests distinct clockwise $I$-$E$ hysteresis as applied field reaches 3.33 MV/cm, unambiguously demonstrating the memristive resistive-switching behavior. The RESET (positive bias) and SET (negative bias) transitions occur distinctly in the ultrathin device, even without requiring an initial forming step.

When ultrathin ferroelectric devices are operated above $E_c$, resistive switching and polarization reversal occur simultaneously, giving rise to a pronounced CPR phenomenon. This interplay profoundly alters the cycling characteristics, yielding anomalous behaviors distinct from conventional $HfO_2$-based ferroelectric capacitors.[11] While conventional $HfO_2$-based devices typically follow a sequential progression of wake-up, stable switching, and eventual fatigue (Figure S3),[11] our ultrathin HZO device demonstrates an unconventional polarization reduction during the initial cycling (Figure 1c-d) process. Figures 1e-1f present the evolution of *P-E* hysteresis loops and corresponding current curves at different cycling stages, respectively. Initially, the ultrathin HZO device exhibits pronounced resistive-switching behavior without a well-defined ferroelectric response, despite the applied field exceeds $E_c$. However, after 100 switching cycles, the maximum current densities under positive and negative biases decreased markedly from 2.3 A/cm² and -4.4 A/cm² to 0.8 A/cm² and -1.3 A/cm², respectively (Figure 1f). This reduction in current density unequivocally confirms the transition of the device from a low-resistance (LRS) to a high-resistance state (HRS). Subsequently, the intensity of the polarization switching current peak shows significant enhancement. Operating in the HRS with suppressed leakage, the device yields "thinner" *P-E* loops (Figure 1e) with pronounced polarization switching behavior (Figure 1f). The apparent polarization reduction observed in Figures 1c-1f underscores the competitive interaction between resistive switching and polarization switching, as their coexistence mutually suppresses independent characterization of each characteristic.

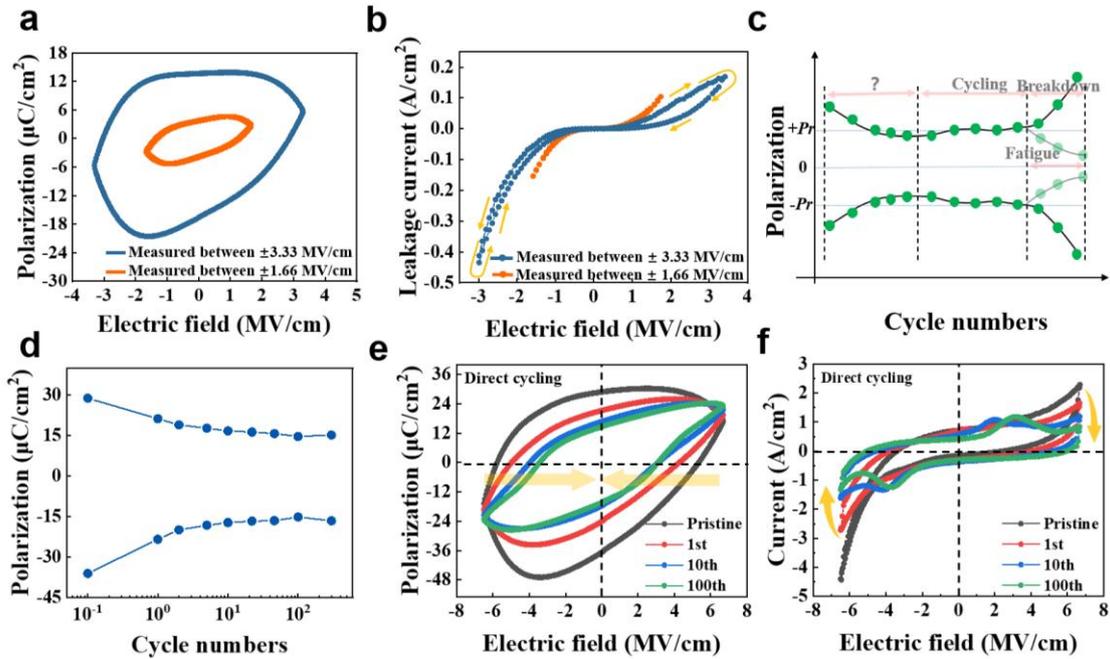

**Figure 1. The electric performances and cycling behavior for the Pt/HZO/LSMO device**. a) *P-E* loops of the pristine device measured beneath $E_c$. b) leakage current-electric field (*I-E*) curves measured within 1.66 MV/cm and 3.33 MV/cm respectively. c) A brief sketch of the whole cycling performance for the ultrathin HZO ferroelectric device. The period labelled by the question mark represents a duration that differs from the initial cycling performances of conventional ferroelectric devices. d) The specific cycling behavior of the ultrathin HZO device at the initial stage. e) *P-E* loops and f) corresponding switching currents for the ultrathin Pt/HZO/LSMO device directly cycled above $E_c$.

To investigate the origin of the lossy characteristic, conductive atomic force microscopy (C-AFM) was utilized to probe the local conductivity of the ultrathin film and elucidate the underlying mechanisms governing its resistive-switching behavior. The pristine HZO film exhibits atomic-scale flatness with a root-mean-square roughness of only ~110 pm (Figure S4).

Current mapping acquired under the -2 V tip voltage reveals spatially inhomogeneous conduction (Figure 2a), indicating localized regions with distinct resistance states. Quantitative analysis of the current distribution (Figure 2c) along the white dotted line (Figure 2a) further confirms multiple LRS regions on the pristine surface. Given the resistive-switching characteristics of the film, the resistance state can be modulated by external electric stimuli. Remarkably, these conductive sites vanish upon applying a +2 V voltage (Figures 2b, 2d), signifying a transition to the HRS. These observations demonstrate that the resistive-switching behavior detected in ultrathin HZO arises from $V_O$ migration and the dynamic formation/disruption of $V_O$ conductive filaments under electric fields (Figure 2e).[32] Besides, $V_O$ in $HfO_2$ films tend to be arranged in a structured and well-ordered manner, resembling chains[7] or planes.[33] Under this circumstance, the probability of ordered $V_O$ filaments extending throughout the entire film is considerably higher for the ultrathin films compared to their thicker counterparts. This facilitates the electric-forming-free resistive-switching behavior observed in the ultrathin HZO devices and accounts for their pronounced lossy dielectric characteristics.

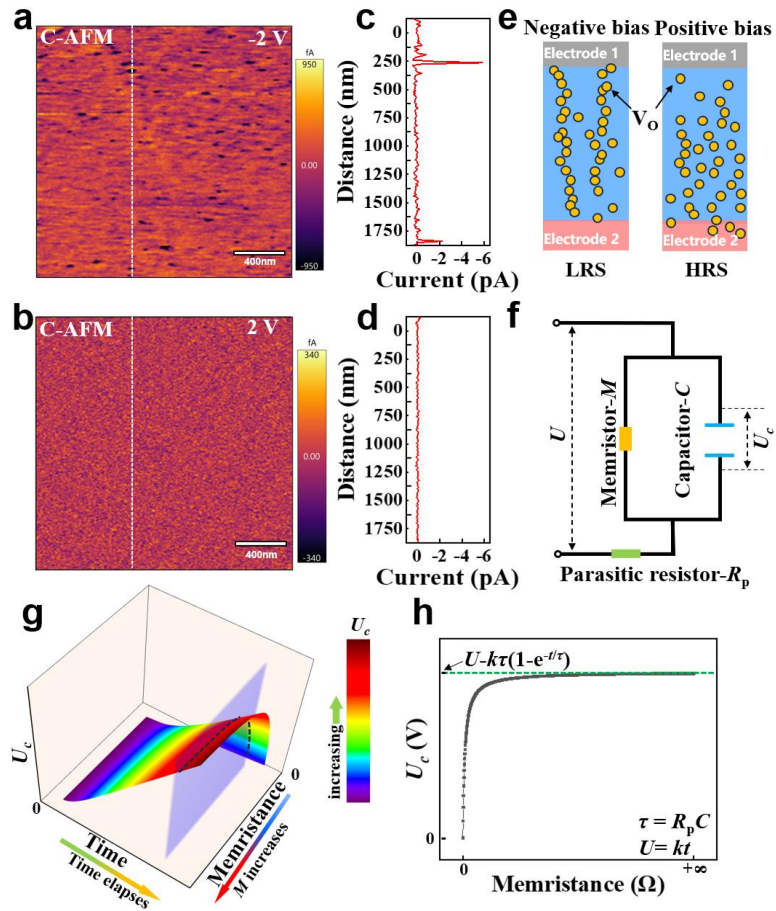

**Figure 2. Mechanisms of the conductive states in the ultrathin HZO device.** Current mapping of the HZO film under (a) a -2 V bias and (b) a +2 V bias. c, d) Line profiles of current distribution along the white dotted lines in (a) and (b), respectively. e) Schematics of $V_O$ distribution conditions within the HZO film under electric fields of different polarities. f) A simplified equivalent circuit model of the Pt/HZO/LSMO device system. g) Relationship among $U_c$, time, and memristance (*M*). h) Extracted $U_c$-*M* curve for a fixed time parameter, derived from the curved surface in (g).

As schematically depicted in Figure 2f, the ultrathin Pt/HZO/LSMO ferroelectric device can be modeled as a parallel circuit comprising a capacitor and a memristor, effectively capturing the intrinsic electrical characteristics of the system. This configuration can be represented as an active two-terminal network (Figure S5a-b), enabling a simplified electrical analysis. According to the Thevenin's theorem,[34,35] the system can be further simplified into an equivalent circuit, as shown in Figure S5c, where the equivalent electromotive force ($E$) and internal resistance ($R_e$) are given by:

$$E = U \frac{M}{R_p + M} \tag{1}$$

$$R_e = \frac{MR_p}{R_p + M} \tag{2}$$

Where $U$, $M$ and $R_p$ are the electric voltage across the system, the memristance and the parasitic resistor value, respectively. Accordingly, it can be calculated by the following equations:

$$iR_e + U_c = E \tag{3}$$

$$R_e C \frac{dU_c}{dt} + U_c = E \tag{4}$$

Where $i$, $C$, $U_c$ and $t$ represent the current of the circuit shown in Figure S5c, the capacitance of the capacitor, the voltage drop across the capacitor and the time elapsed since the load of the voltage, respectively. Furthermore, regarding the voltage drop, it should be noted that the polarization reversal occurs once $U_c$, rather than $U$, reaches the coercive force threshold of the device.

When the pulse waveform (Figure S6) is applied to the system, the voltage drop $U_c$ across the ferroelectric capacitor part can be described as:

$$U_c = \frac{kMt}{R_p + M} - \frac{kR_p M^2 C}{(R_p + M)^2}\left(1 - e^{\frac{t(R_p + M)}{MR_p C}}\right) \tag{5}$$

Where, $k$ is the constant representing the changing rate of the applied driving force (Figure S6). As visualized in Figure 2g, which illustrates formula (5), $U_c$ varies with different values of $M$ over time. For clarity, we extract a time-fixed cross-section from the 3D plot (Figure 2g) and present it in Figure 2h, which reveals a monotonic positive correlation between $U_c$ and $M$. Hence, a higher memristance $M$ leads to a greater voltage drop $U_c$ across the capacitor. Conversely, a smaller $M$ results in a lower $U_c$, which fails to meet the coercive force required for polarization reversal, thereby contributing to the initially observed lossy attribute and faint switching peaks in Figure 1e and Figure 1f.

**Effects of the CPR phenomenon on electric characterizations**

The electric-field-driven migration of $V_O$ is an inherent and inevitable process, directly influencing the resistive-switching behavior, which is intrinsically coupled with the polarization reversal in $HfO_2$-based ferroelectric thin films. Therefore, beyond the aforementioned anomalous cycling behavior, the CPR phenomenon can significantly impact the ferroelectric switching characteristics of the ultrathin device. The DC current-field curves (Figure 1b) reveal the evolution of memristance during polarization reversal. Figure 3a presents a snapshot of the memristance state throughout this process, allowing its variation to be systematically traced as a function of the applied voltage. It can be inferred that the device transitions from a HRS to a LRS as the electric field shifts from positive to negative bias, altering $U_c$ accordingly through the resistive-switching process. As illustrated in Figure 3a and 3b, point I marks the onset of polarization switching. Despite points I and II experience identical applied voltages (i.e., $|U^{II}|$ = $|U^{I}|$), the resultant capacitor voltage drop differs substantially ($|U_c^{II}|$ < $|U_c^{I}|$) due to reduced memristance ($M^{II}$ < $M^{I}$), a direct consequence of resistive-switching effect. Therefore, according to Equation (5), the memristance decrease necessitates both extended switching duration (t) and elevated driving amplitude ($U = kt$) to achieve the coercive threshold ($U_c$), with position II' marking this critical transition point. Consequently, polarization reversal under the negative bias requires a higher field. Similarly, unlike ideal insulating ferroelectrics, the pronounced lossy behavior and changed $M$ values induced by the CPR effect substantially prolong the switching time for both polarization states, thereby increasing $E_c$. Additionally, as illustrated in Figure 3b, the leakage current, which exhibits resistive-switching behavior, also displays hysteresis characteristics, further complicating the accurate assessment of ferroelectric switching performances. It couples with the intrinsic polarization switching current, converting the ideal square-like polarization switching current loop into a broader, tilted one. As a result, extrinsic contributions derived from the resistive-switching effect distort the polarization hysteresis loops and inflate the apparent polarization values, a prevalent issue in the characterization of ultrathin ferroelectrics. It is important to note that, even the mere drift of $V_O$ can also generate a ferroelectric-switching-like current response,[36] further obscuring the distinction between intrinsic and extrinsic ferroelectricity in ultrathin devices.[37]

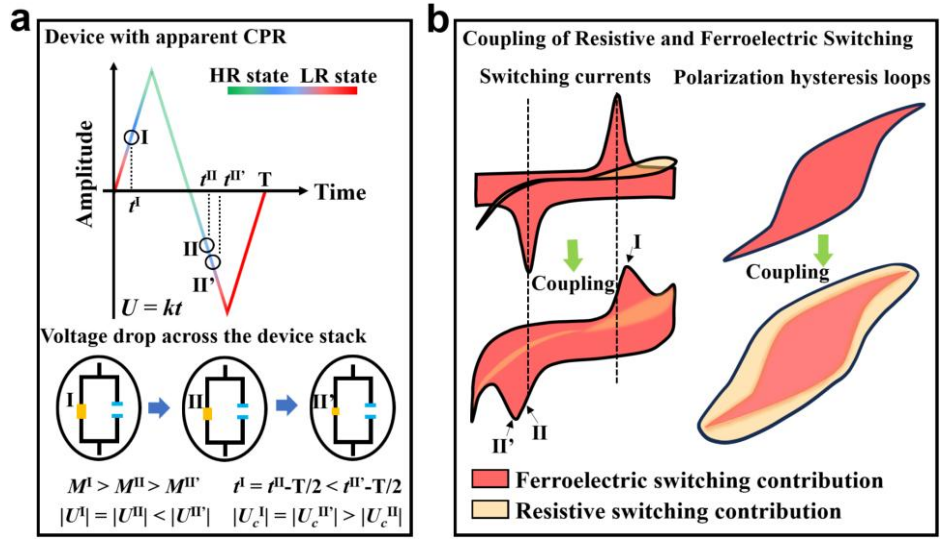

**Figure 3. Schematic illustration of the impact of the CPR phenomenon on the switching characteristics of the ultrathin HZO ferroelectric device.** a) The snapshot of the memristance change and the variation in $U$ and $U_c$ during the polarization reversal process. T is the time of an electric pulse period and $t^I$, $t^{II}$, $t^{II'}$ represent the time of the marked points I, II and II' respectively. Additionally, the $M^x$, $U^x$ and $U_c^x$ represent the specific memristance, voltage applied to the circle and the voltage drop across the capacitor at the marked point $x$ ($x$ = I, II and II') respectively. b) Coupling effects of the resistive and ferroelectric switching. The red regions stand for the ferroelectric switching contributions and the yellow places symbolize the resistive switching contributions.

The intrinsic ferroelectric properties of ultrathin HZO devices could be obscured due to their coupling with the resistive-switching behavior. The Positive-Up-Negative-Down (PUND) method has been widely employed to subtract dielectric and leakage contributions during polarization switching, enabling the extraction of intrinsic ferroelectric characteristics.[38] However, its effectiveness is compromised in ferroelectric systems with CPR effect. The pronounced hysteresis in resistive-switching current curves prevents the complete elimination of extrinsic contributions through conventional PUND measurements, potentially leading to misinterpretations — most notably, inflated polarization values. As illustrated in Figure 4a, to investigate the variation in PUND responses across different memristive states, we employed a specially designed 'PUNDPU' pulse sequence. This sequence combines a standard PUND (pulses 1-4 with a pulse duration of 100 μs), an equivalent NDPU (pulses 3-6 with a pulse duration of 100 μs), and a shorter write pulse (with a pulse duration of 50 μs). Due to varying electrical histories applied before points i (after a shorter SET write pulse) and ii (following two prolonged repeated SET 'ND' pulses), the device exhibits different memristive states, enabling the analysis of their impact on polarization switching. As shown in Figure 4b, the $U_c$ at position ii is smaller than that at position i. Therefore, the current response to pulse-5 exhibits a delayed switching peak, along with a pronounced contribution from the resistive-switching hysteresis within the peak region, in stark contrast to the current response observed for pulse-1 (Figure 4c). Notably, this additional resistive-switching contribution cannot be eliminated by the Up current associated with pulse-6 due to its hysteresis characteristic. Consequently, as exemplified in Figure S7, the CPR phenomenon causes the same device to exhibit varying polarization values when measured using the "PUND" method.

It is worth mentioning that a higher applied field can exacerbate the CPR effect, introducing more significant artificial contributions to the polarization values. As shown in Figure 4d, prior to performing the standard PUND measurement with an 8.33 MV/cm peak field, the device was subjected to different set amplitudes to establish distinct resistance states. As a result, after being programmed at -8.33 MV/cm, the total switching current of the device contains a significantly enhanced contribution from the resistive switching, accompanied by a delayed switching current peak (Figure 4e). This leads to a noticeable increase in the nominal polarization value and a higher $E_c$ (Figure 4f). Notably, the contribution of the resistive-switching effects can be further amplified under higher electric fields, provided the device remains functional (i.e., without breakdown). For instance, under a PUND measurement with ±10 MV/cm and a -10 MV/cm SET amplitude, the nominal remnant polarization ($2P_r$) of the device reaches 50.5 μC/cm$^2$ (Figure S8), representing a 189% enhancement compared to its pristine state (Figure 4g). This phenomenon not only provides critical insights into the extrinsic origins underlying the large discrepancy between nominal $P_r$ value obtained via conventional PUND technique (34 μC/cm$^2$) and intrinsic $P_r$ value (< 9 μC/cm$^2$) obtained from atomic displacements, as documented in prior investigations,[6,22] but also accounts for the exaggerated $P_r$ value of up to 387 μC/cm$^2$ observed in the HZO film.[25] To further verify the role of $V_O$ drift, the Pt/HZO/LSMO device was characterized at liquid-nitrogen temperature (77 K), where $V_O$ diffusion is significantly suppressed. The resulting PUND curves (Figure 4h-i) showed remarkable consistency across different set voltages. This further confirms the influence of CPR on the polarization switching behaviors of HZO ultrathin devices.

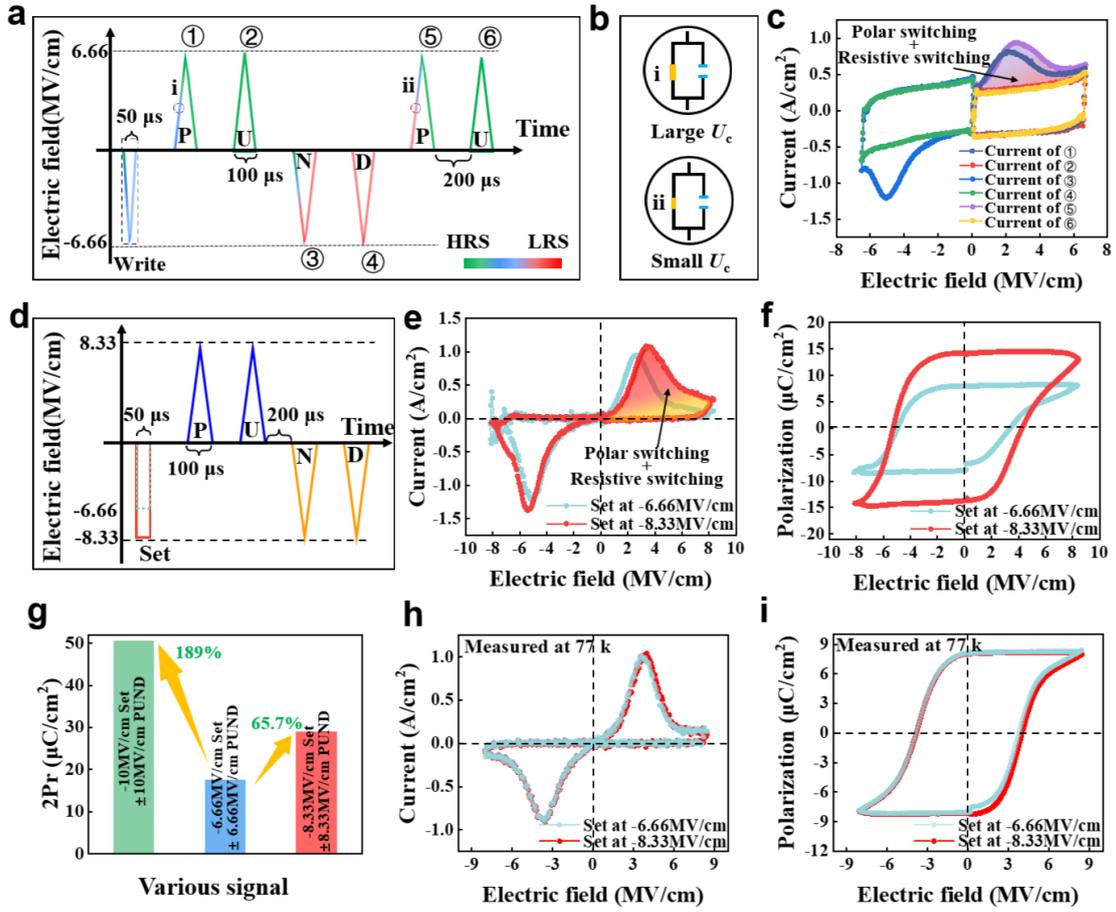

**Figure 4. Effects of CPR on the polarization switching performances of the Pt/HZO/LSMO device.** a) The schematic of the design PUNDPU pulse sequence (numbered 1-6), the change of resistance conditions along with the applied electric fields. b) The $U_c$ conditions of the device located at i and ii points marked in the (a). c) Corresponding current curves generated by the PUNDPU pulses shown in (a), and the signals labeled 1–6 correspond to the current responses generated by electrical pulses 1–6 in panel (a), respectively. d) The designed PUND pulse with various SET amplitudes. e) The switching currents and f) corresponding PUND curves of the same device measured with different SET signals. g) The change of nominal remnant polarization values with different SET and read pulses. h) The switching currents and (i) corresponding PUND loops measured with pulses shown in (d) at 77 k.

**Suppressing resistive-switching via $V_O$ migration control in ultrathin HZO films**

The CPR phenomenon introduces artificial contributions to ferroelectric characterizations, increasing $E_c$ and obscuring accurate assessment of the material's intrinsic properties. Therefore, as discussed, the strong interplay between resistive-switching and polarization reversal necessitates effective mitigation strategies for the resistive-switching behavior in ultrathin HZO devices. Inspired by the unusual cycling behavior observed in Figure 1e and the pronounced migration of $V_O$ in ultrathin films (Figure 2a-2b), it can be inferred that the resistance state of the ultrathin films can be modulated through $V_O$ migration. In thicker films, $V_O$ conductive filaments are typically formed via an electric-forming process induced by a DC pulse (Figure 5a). By contrast, in ultrathin films, tuning the magnitude and frequency of an alternating electric field continuously reorients $V_O$, resembling an effective "shaking" motion. This process causes continuous $V_O$ filaments to rupture at weak points and disperse into isolated clusters, ultimately restoring the high-resistance state of the high-$\kappa$ film. This process, referred to the dielectric training (D-training), offers a promising approach to mitigating resistive-switching effects.

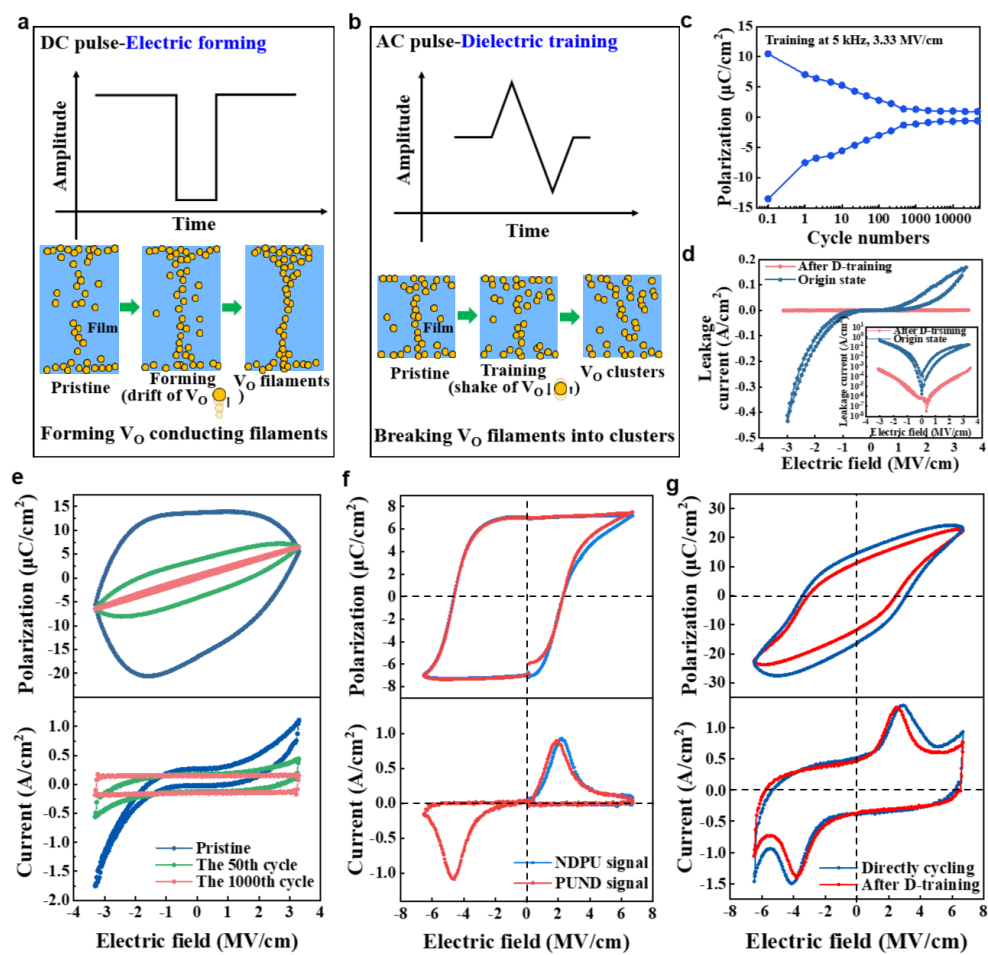

**Figure 5. Illustrations of dielectric-training process.** Visualizations of the (a) electric-forming and (b) dielectric-training processes, with accompanying depictions of $V_O$ migration under each condition. c) The dielectric training process for the 6 nm-thick Pt/HZO/LSMO device, showing the changing trend of the nominal polarizations under the 3.33 MV/cm, 5k Hz AC signal. d) The leakage currents for the device measured before and after the dielectric training treatment. The figure inset shows currents with a logarithmic scale. e) The nominal *P-E* loops and corresponding switching currents of the ultrathin HZO device during the D-training process. f) The "PUNDPU" measurement results and (g) *P-E* loops for the trained device measured above $E_c$, with their corresponding switching currents shown alongside.

To optimize the D-training process, various parameters were systematically explored. Figure S9 demonstrates that AC pulses across different frequencies and amplitudes can effectively facilitate D-training by driving $V_O$ migration. However, higher voltages or lower frequencies would increase the risk of breakdown in ultrathin devices, a phenomenon that aligns with the time-dependent-dielectric-breakdown (TDDB) model.[39,40] To sum up, the selection of electrical signals plays a crucial role in determining the arrangement and distribution of defects within HZO thin films (Figure S10), which in turn dictates the electrical performance of HZO-based devices, encompassing aspects such as leakage, dielectric response, split-up behavior,[41] breakdown, and endurance. Meanwhile, the thickness of HZO films also affects the training effect (Figure S11). For the 2.5 nm film, it demonstrates an enhanced probability for $V_O$ filaments extending through the entire film. Also, the ultrathin thickness favors the tunneling current. Both contribute to higher nominal polarizations and a prolonged training process in the 2.5 nm ultrathin device. Conversely, $V_O$ filaments within the 10-nm-thick HZO film show a reduced likelihood of connecting both interfaces, yielding a rapid training completion. Notably, due to the metastable nature of $HfO_2$-based ferroelectrics[28] and inherent process variations, the thickness-dependent training process may be varying across different reports.[25]

In light of this, a 3.33 MV/cm (below $E_c$) pulse at 5 kHz was used to train the 6-nm-thick HZO film (Figure 5c), resulting in a three-order-of-magnitude reduction in leakage current and significant suppression of resistive-switching behavior (Figure 5d). This treatment effectively mitigated the CPR effect, enabling the film to recover its dielectric nature (Figure 5e). Moreover, the D-trained device was also characterized using the "PUNDPU" measurement, which follows a signal profile analogous to that shown in Figure 4a. Compared with the untrained counterparts (Figure S7), the PUND and NDPU responses of the D-trained sample (Figure 5f) exhibit better consistency, similar polarization values, and narrower hysteresis loops. Besides, as shown in Figure 5g, the D-trained sample exhibits better ferroelectric characteristics, including reduced leakage current, minimized extrinsic polarization contributions, and a lower $E_c$. In summary, the D-training process not only reveals pronounced $V_O$ migration in ultrathin ferroelectric devices, but also provides a comprehensive understanding of the substantially amplified impact of defect migration on electric performances—including resistive-switching characteristics, dielectric properties, and ferroelectric behaviors—in the ultrathin thickness condition.

Figure 6a compares the endurance performance of devices subjected to D-training and direct cycling (characterized using the pulse waveform shown in Figure S12a), respectively. The D-trained sample exhibits a sequential transition from stable cycling to a leaky state and eventually to fatigue. For the directly cycled device, Figure S12b provides an overall view of its cycling behavior. It is noteworthy that the device escapes from the breakdown behavior into the fatigue situation. Meanwhile, apart from the breakdown duration between the leaky and fatigue stages, it essentially shares the same cycling experience with the trained sample. This suggests that the D-training process effectively disrupts continuous $V_O$ filaments, which helps improve breakdown strength. Unexpectedly, both devices ultimately converged to the same final state, exhibiting dielectric characteristics in the fatigue region (Figure S12c).

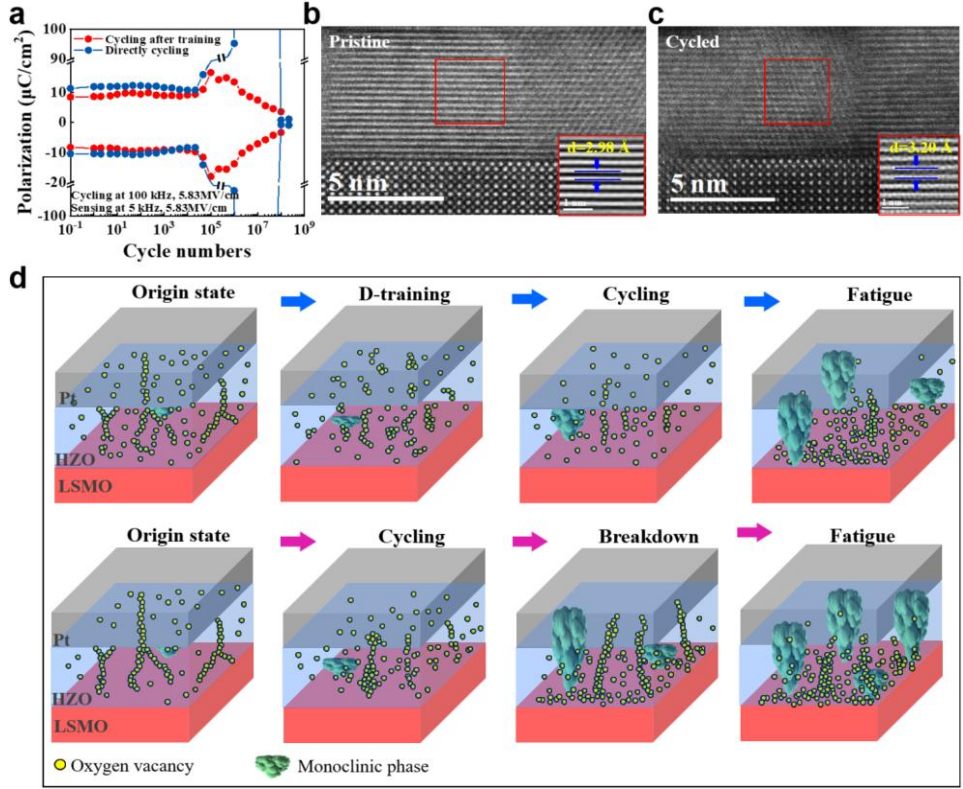

**Figure 6. Demonstration of the influence of CPR on cycling performances of the Pt/HZO/LSMO devices.** a) Endurance performance for samples subjected to D-training and direct cycling, respectively. b-c) the HAADF-STEM images for the sample before and after the endurance measurement, with the inset figures demonstrating the variation of *d*-spacings before and after the endurance measurement. d) Schematics for the migration trails of $V_O$ within the HZO device under different electric measurement.

As discussed above, due to the CPR effect, the device exhibits variability in its memristance states during polarization switching. Similarly, under repeated rectangular endurance pulses, the $U_c$ across the ferroelectric capacitor during the endurance measurement can be described as (Figure S13):

$$U_c = E - Ee^{-\frac{t}{R_eC}} = U\frac{M}{R_p+M}(1 - e^{-\frac{t(R_p+M)}{MR_pC}}) \qquad (6)$$

Consequently, based on the memristive characteristic of the device, the *M* at 5.83 MV/cm is larger than that at -5.83 MV/cm, leading to a higher $|U_c|$ under positive bias. As illustrated in Figure S13c, the variation in $|U_c|$ induces asymmetrical $V_O$ migration velocities under opposite bias polarities per cycle, resulting in a net drift of $V_O$ towards the bottom electrode over multiple cycles. The oriented migration of $V_O$ during endurance testing ultimately leads to the complete rupture of $V_O$ filaments and the progressive oxidation of HZO near the Pt/HZO interface region. This transition shifts the devices from a leaky or breakdown state to a predominantly dielectric state. Meanwhile, the oxygenated Pt/HZO interface region can no longer sustain the oxygen-deficient environment required to stabilize the polar phase of HZO, potentially inducing a phase transition from the metastable ferroelectric structure to the energetically favored monoclinic paraelectric phase, as evidenced by the increased interplanar spacing[42] after cycling (Figures 6b-c). In addition, micro-area X-ray diffraction (Figures S15a–S15c) was employed to track the macroscopic structural evolution of the ultrathin HZO device in its initial, D-trained, and fatigued states (Figures S15d-S15f).[43] The structural profiles remain nearly identical between the initial and D-trained states, whereas a pronounced increase in the paraelectric phase fraction is observed in the fatigued situation. These results corroborate the micro-structural findings observed by STEM. This structural transformation accounts for the reduction in polarization in the fatigue stage. As schematically depicted in Figure 6d, based on the cycling behaviors of the device, the migration trajectory of $V_O$ at different cycling stages can be traced within the ultrathin HZO film. As shown, different applied fields lead to distinct migration dynamics, directional displacement, and diverse distribution of $V_O$, which in turn alter the electrical response of the device. The D-training process effectively disrupts conductive $V_O$ pathways, thereby reducing leakage current and improving cycling endurance. In contrast, the direct cycling process induces stronger directional $V_O$ migration, resulting in less effective leakage suppression and higher vulnerability to breakdown and fatigue. In this context, the CPR phenomenon fundamentally reflects the interplay between polarization switching and oxygen defect migration. Although D-training mitigates resistive-switching effects, it cannot eliminate defect migration under electrical field. Consequently, polarization fatigue remains inevitable in the Pt/HZO/LSMO device over extended cycling.

**Mitigated resistive-switching and improved ferroelectric performances through the symmetric interface**

The competitive interplay between polarization reversal and resistive switching would result in exaggerated polarization values, increased $E_c$, and degraded device reliability, representing a critical bottleneck in developing high-performance ultrathin $HfO_2$-based ferroelectric devices. Apart from the D-training method, we mitigate the resistive-switching behavior from multiple perspectives. Previous studies have demonstrated that oxide films sandwiched between asymmetric electrodes with different work functions, such as the Pt/HZO/LSMO heterostructures, tend to exhibit more pronounced resistive-switching behaviors.[44–46] To address this, the Pt top electrode has been replaced with an LSMO layer to construct the symmetric LSMO/HZO/LSMO device, thereby increasing both the SET and RESET threshold voltages within the device system. Meanwhile, considering the mobile nature of oxygen defects in $HfO_2$-based films, LSMO symmetric oxide electrodes can serve as oxygen reservoirs,[43,47] enabling the reversible acceptance and donation of $V_O$. This helps alleviate the accumulation of defects like $V_O$ at the HZO-electrode interface.[48–50]

As shown in Figure 7a, when subjected to the same pulse signal shown in Figure 4a, the PUND and NDPU curves for the LSMO/HZO/LSMO device are substantially identical, indicating effective suppression of resistive-switching behavior. This diminished CPR effect markedly improves polarization efficiency, yielding a lower $E_c$ in the symmetric capacitor. The synergistic combination of suppressed resistive-switching, symmetric architecture, and oxide electrodes effectively restricts directional $V_O$ migration and prevents oxygen defect accumulation at the HZO-electrode interfaces. This leads to impressive endurance performance (Figure 7b), characterized by mild fatigue behavior and remarkable endurance life exceeding $10^{10}$ cycles. Moreover, leakage current (Figure 7c) in the symmetric device is significantly reduced compared to the Pt/HZO/LSMO device, with no discernible resistive-switching behavior. Furthermore, synchrotron micro-area XRD studies were performed to probe the phase transformation of HZO before and after the endurance test. Benefiting from the design outlined above, the ferroelectric orthorhombic phase remains extremely stable, with only a slight increase in the monoclinic phase signature after $10^8$ cycles (Figure 7d-e), which explains the exceptional endurance from a phase-stability perspective. Similarly, $V_O$ migration behavior can be inferred from the switching characteristics. As shown in Figure 7f, the device with symmetric oxide electrodes shows a more homogeneous distribution of $V_O$ and less segregation of oxygen defects during cycling compared to the Pt/HZO/LSMO device.

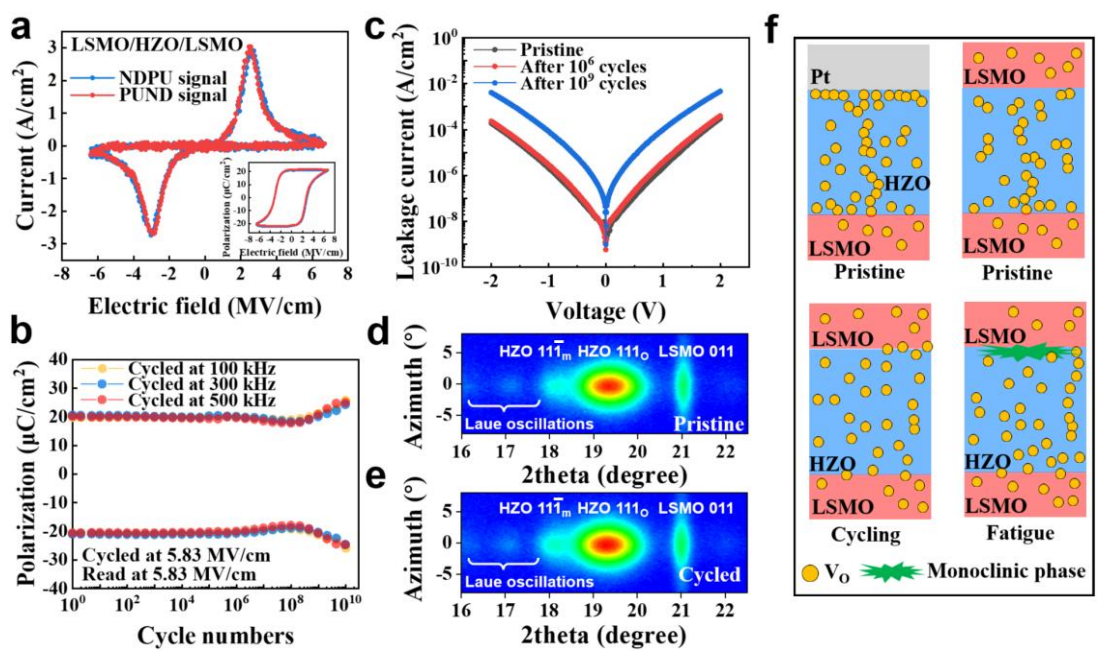

**Figure 7. The switching characteristics for the HZO-based device with symmetric LSMO electrodes.** a) The switching currents and corresponding hysteresis loops measured with PUND and NDPU signals. b) The endurance performances of the LSMO/HZO/LSMO device, measured across various frequencies, exhibit a high degree of consistency. c) Leakage currents for the LSMO/HZO/LSMO device. d-e) The micro-area hard X-ray diffraction patterns for the LSMO/HZO/LSMO sample before and after the endurance test, measured at the BL15U1 beamline station in Shanghai Synchrotron Radiation Facility (SSRF) with a wavelength of 1 Å. f) The description of $V_O$ movement condition in the Pt/HZO/LSMO and LSMO/HZO/LSMO devices at different stages.

**DISCUSSION**

Our study thoroughly investigates the resistive-switching behavior and its intricate relationship with polarization switching performance in ultrathin HZO ferroelectric device. A simplified model is developed to delineate the resistive-switching process and the alterations in memristance states during electrical switching. By which means, we have successfully decoupled the intrinsic polarization switching from the resistive switching. Our findings further reveal that resistive-switching behavior is closely linked to more pronounced $V_O$ migration behavior in ultrathin films, leading to reduced polarization efficiency, exaggerated polarization values, increased $E_c$, and inferior cycling stability. In view of this, a symmetric HZO-based device, sandwiched between two LSMO electrodes, is fabricated to suppress resistive-switching behavior in the ultrathin HZO film. This configuration mitigates the side effects of the CPR phenomenon, resulting in significantly improved performance, including reduced $E_c$, enhanced phase stability, and superior endurance. Our findings offer valuable insights in the abnormally high polarization values and poor cycling performances observed in ultrathin ferroelectric devices and provide a feasible solution to alleviate these issues.

**METHODS**

**Sample fabrication**: HZO ferroelectric films were prepared by pulsed laser deposition (PLD) method using a KrF excimer laser with a wavelength of 248 nm. The substrate and bottom electrode are SrTiO$_3$ (001) single crystal (STO) and La$_{0.67}$Sr$_{0.33}$MnO$_3$ (LSMO) respectively. Among them, the LSMO bottom electrode was grown at 700 ℃ surrounded by 20 Pa O$_2$ atmosphere and the laser fluence and frequency were 0.85 J/cm$^2$ and 3 HZ. After the deposition of LSMO, the temperature was cooled to 600 ℃ and O$_2$ pressure decreased to 15 Pa for the preparation of HZO films. the sputtering energy density and frequency for HZO are 1.35 J/cm$^2$ and 2 HZ respectively. As for the LSMO/HZO/LSMO symmetric architecture, the top LSMO layer was deposited under the 20 Pa O$_2$ atmosphere at 600 ℃. The laser fluence and frequency used for the top electrode deposition were 0.85 J/cm$^2$ and 5 HZ. The metal-ferroelectric-metal (MFM) devices of HZO with 100 nm Pt as top electrodes were fabricated by magnetron sputtering after the patterning by photolithography.

**Structural characterizations**: The structure and thickness of HZO films were examined by X-ray diffraction (XRD) and X-ray reflection (XRR) using a Rigaku-Smartlab (9 kW) diffractometer with Cu K$_{\alpha 1}$ radiation. The micro-area X-ray diffraction was measured at the BL15U1 beamline station in Shanghai Synchrotron Radiation Facility (SSRF). The wavelength and diameter of the X-ray is 1 Å and 5 μm respectively. The obtained data was analyzed using the *Dioptas* program[51]. Samples for STEM observations were prepared by the focused-ion-beam (FIB) lift-out method. The C and W coating were deposited on the devices to protect the surface of devices prior to the cutting process. Firstly, the voltage of 30 kV and current of 0.44 nA for the ion beam were used to polish the TEM samples. Afterwards, the current was gradually reduced to 41 pA. Finally, the voltage of 5 kV, 2 kV and 1 kV for ion beam were used to clean the surface amorphous. Cross-section view of HAADF-STEM images were acquired by a double-aberration-corrected scanning transmission electron microscope (Spectra 300, ThermoFisher Scientific) equipped with a monochromator, a Gatan 1069 EELS system and K3 camera operated at 300 kV.

**Electric performance characterization**: the Oxford instrument MFP-3D Infinity was used to detect the surface morphology and conductive state of the prepared sample. Specifically, the C-AFM measurement was performed directly on the bare HZO surface without top electrode, but platinum-coated conductive AFM probes were used to ensure reliable electrical contact.

The electrical properties of the HZO ferroelectric films—including polarization–electric field ($P$–$E$) hysteresis loops, pulse-based measurements (PUND and other custom pulse sequences), leakage current, and endurance characteristics—were characterized using an aixacct TF3000 analyzer in the MFM capacitor configuration with a top electrode diameter of 25 μm. For PUND and other custom pulse sequences, the time interval between consecutive pulses is set to 200 μs. The signals used for the $P$-$E$ test and the pulse measurement are 5 kHz triangle pulses. The endurance tests were cycled with the 100 kHz rectangle pulse and sensed with the 5 kHz triangle signal. During the electrical test, Pt electrodes or top electrodes were connected to the positive bias while the LSMO bottom electrodes was grounded.

## RESOURCE AVAILABILITY

### *LEAD CONTACT*

Further information and requests for resources and reagents should be directed to and will be fulfilled by the lead contact, Zuhuang Chen (zuhuang@hit.edu.cn).

### *MATERIALS AVAILABILITY*

This study did not generate new unique reagents.

### *DATA AND CODE AVAILABILITY*

All data reported in this paper will be shared by the lead contact upon request

## ACKNOWLEDGMENTS

This work was supported by National Natural Science Foundation of China (Grant Nos. 92477129 and 52525209) and the Guangdong Basic and Applied Basic Research Foundation (Grant No. 2024B1515120010). Z.H.C. acknowledges the financial support for Outstanding scientific and technological innovation Talents Training Fund in Shenzhen. S.Q.D. acknowledges the financial support from National Natural Science Foundation of China (Grant No. 22375015), the Open Research Fund of the Guangxi Key Laboratory of Information Materials (Grant No. 221031-K) and the Interdisciplinary Research Project for Young Teachers of USTB (Fundamental Research Funds for the Central Universities) (Grant No. FRF-IDRY-23-006). S.L. has been supported by National Key R&D Program of China (Grant No. 2021YFA1202100) and National Natural Science Foundation of China (Grant No. 12361141821). H.L.W. has been supported by National Natural Science Foundation of China (Grant No. U24A2013). The authors thank the staff from Shanghai Synchrotron Radiation Facility (SSRF) at BL15U1.

## AUTHOR CONTRIBUTIONS

Z.H.C. and C.Z. conceived and designed the research. C.Z. fabricated the films and carried out electric measurements with the assistance of Y.Y.S. and H.L.H. C.Z. performed the XRD measurements with the assistance of H.L.H., S.Z.H, and Y.Y.S. Z.Q.R performed the C-AFM measurement. J.Y.Z and S.L. conducted the model of the testing circle. C.Z., S.Q.D., S.D., and Z.H.C. wrote the manuscript. Z.H.C. supervised this study. All authors discussed the results and commented the manuscript.

**DECLARATION OF INTERESTS**

The authors declare no conflict of interest.

**SUPPLEMENTAL INFORMATION**

**Figure S1-S15**